# Fractal Relativity, Generalized Noether Theorem and New Research of Space-Time

*Yi-Fang Chang*
*Department of Physics, Yunnan University, Kunming 650091, China*
(e-mail: yifangchang1030@hotmail.com)

**Abstract**

First, let the fractal dimension D=n(integer)+d(decimal), so the fractal dimensional matrix was represented by a usual matrix adds a special decimal row (column). We researched that mathematics, for example, the fractal dimensional linear algebra, and physics may be developed to fractal and the complex dimension extended from fractal. From this the fractal relativity is discussed, which connects with self-similarity Universe and the extensive quantum theory. The space dimension has been extended from real number to superreal and complex number. Combining the quaternion, etc., the high dimensional time $ict \to ic_1 t_1 + jc_2 t_2 + kc_3 t_3$ is introduced. Such the vector and irreversibility of time are derived. Then the fractal dimensional time is obtained, and space and time possess completely symmetry. It may be constructed preliminarily that the higher dimensional, fractal, complex and supercomplex space-time theory covers all. We propose a generalized Noether theorem, and irreversibility of time should correspond to non-conservation of a certain quantity. Resumed reversibility of time and possible decrease of entropy are discussed. Finally, we obtain the quantitative relations between energy-mass and space-time, which is consistent with the space-time uncertainty relation in string theory.

## 1. Fractal and Its Development

1975 B.Mandelbrot [1,2] proposed and studied "fractal" whose form is extremely irregular and/or fragmented at all scales. We think that fractal possesses two basic characters: the self-similarity for different scales and the fractal dimension, which may be non-integer (fractional or irrational number). Therefore, we extended the fractal dimension D into the complex dimension in both aspects of mathematics and physics [3], whose representation is:

$$D_z = D + iT . \quad (1)$$

When the complex dimension is combined with relativity, whose dimensions are three real spaces and an imaginary time. It may express a change of the fractal dimension with time or energy, etc., and exists in the fractals description of meteorology, seismology, medicine and the structure of particle, etc [4].

Further, fractal dimension may be extended to superreal and supercomplex dimension [5]. We researched many aspects of development of fractal dimension in mathematics and physics [5]. Assume that the fractal dimension D=n(integer)+d(decimal). From this the fractal dimensional linear algebra, analysis and set, etc., may be developed. For example, the fractal dimensional matrix was that a usual matrix adds a special decimal row (column), whose D dimensional square

matrix is defined as:

$$\begin{pmatrix} a_{11} & ... & a_{1n} & a_{1D} \\ ... & ... & ... & ... \\ a_{n1} & ... & a_{nn} & a_{nD} \\ a_{D1} & & a_{Dn} & a_{DD} \end{pmatrix}, \qquad (2)$$

in which the final row and column should be understood as a special decimal dimension d. So all of linear algebra can be applied by the same way, but difference is only final row and column [5]. The fractal analysis relates to differential, integral and operational calculus of non-integer order. The integral operator $l$ is extended to any positive number power as:

$$l^{\lambda} = \left\{ \frac{t^{\lambda-1}}{\Gamma(\lambda)} \right\}. \qquad (3)$$

$\lambda \to -\lambda$ is the differential operator $s=1/l$. Using the same method, we extended vector, tensor, etc., to fractal dimension, even extended various regions of mathematics to the complex number dimension.

Based on the fractal mathematical development, we may research that the fractal dimension and its development are applied to mechanics, statistics and electrodynamics, etc. The fractal quantum theory and its meaning in the particle physics were discussed [5].

## 2. Fractal Relativity

The basic mathematical representation of the special relativity is invariance of the interval $s^2$. Using a method of linear algebra this is invariance of the standard quadric form under a linear transformation. This linear transformation of the interval $s^2$ is:

$$x_{\mu}' = \alpha_{\mu\nu} x_{\nu}. \qquad (4)$$

It is the Lorentz transformation (LT) of v<c for the timelike interval, or the generalized Lorentz transformation (GLT) of $\bar{v}$ >c for the spacelike interval [4,6].

First, we discuss the fractal special relativity with D=n+d dimensional space and a dimensional time. Let $x_4 = ict$, LT of moving speed along any direction [7] in 2+d-dimensional space is:

$$\alpha_{\mu\nu} = \begin{pmatrix} 1+v_x^2 R/v^2 & v_x v_y R/v^2 & v_x v_d R/v^2 & iv_x \gamma/c \\ v_y v_x R/v^2 & 1+v_y^2 R/v^2 & v_y v_d R/v^2 & iv_y \gamma/c \\ v_d v_x R/v^2 & v_d v_y R/v^2 & 1+v_d^2 R/v^2 & iv_d \gamma/c \\ -iv_x \gamma/c & -iv_y \gamma/c & -iv_d \gamma/c & \gamma \end{pmatrix}, \qquad (5)$$

in which $\gamma = 1/\sqrt{1-(v/c)^2}$, $R = \gamma - 1$ and $v_d$ is the speed in decimal dimensional space.

Next, if $v_d = 0$, it is the same with usual LT, only $v_d' = v_d$. If $v_x = v_y = 0$, $v_d = $v, in d+1

dimensional space-time,

$$\alpha_{\mu\nu} = \gamma \begin{pmatrix} 1 & iv/c \\ -iv/c & 1 \end{pmatrix}, \tag{6}$$

i.e., $\quad x_d' = \gamma(x_d - vt), t' = \gamma(t - vx_d/c^2)$. (7)

This is LT of $v_x \to v_d$. Moreover, we may derive the corresponding GLT of $v_d = \bar{v} > c$:

$$x_d' = \bar{\gamma}(x_d - c^2 t/\bar{v}), t' = \bar{\gamma}(t - x_d/\bar{v}), (\bar{\gamma} = 1/\sqrt{1-(c/\bar{v})^2}), \tag{8}$$

In general relativity, if dimension of space is D, so the interval is:

$$ds^2 = \sum_{\mu,\nu=1}^{D+1} g_{\mu\nu} x_\mu x_\nu . \tag{9}$$

Corresponding quantities and formulas should be extended. This connects with some fractal structures of the universe [8,9]. For instance, these distributions on galaxies and clusters of galaxies and superclusters have similarity; the two jet streams on the ultracompact star and the ratio quasars, which are a few million times larger than the former, have also similarity. We found that the average distances (the Titius-Bode law) between the Sun and planets may be represented as a new form [10,11]:

$$r_n = an^2, \tag{10}$$

in which $n$ is integer and $a$ is two constants, respectively, for terrestrial and Jovian planets. It is similar completely with the Bohr atom model. From this we obtained the quantum constants $H = (aGM_O)^{1/2}$ of the solar system, and corresponding quantum theory. Such many quantities of the solar system can be quantized. Further, we derived the astronomical Schrodinger equation, and the distance law is a statistical result of planet evolution. This should be the extensive quantum theory, which has different quantum constants but similar formulations. It is namely a universal quantum theory.

We compare quantitatively the two similar regions: the solar system and the atomic structure, and found that the two middle values of geometrical series are *2.8136m* for distance, and *57.678kg* for mass. Both are about the height of a mankind house and about human weight, respectively [11]. This not only corresponds to the anthropic principle [12], but also is exact. Combining the astronomical quantum theory, the nature shows a self-similar embedded structure.

## 3. High Dimensional Time, Fractal Time and Irreversibility of Time

In relativity the three-dimensional space is real number, and one-dimensional time is imaginary number $x_4 = ict$, so that a unification of space-time corresponds to complex number. When complex number is developed to the quaternion (one of the supercomplex), time should be a three dimensional form:

$$ict \to ic_1 t_1 + jc_2 t_2 + kc_3 t_3. \tag{11}$$

In the extensive special relativity only by c→$c_h$, another invariant velocity $c_h$ may be different velocities, which construct a new variety of space-time systems [4]. This is connected possibly with the many universe proposed by Everett [13], and after many worlds [14]. In Eq.(11) $c_i$ connect possibly with different velocities. When $c_1 = c_2 = c_3 = c$, Eq.(11) is simplified as:

$$ict \to cT = c(it_1 + jt_2 + kt_3). \tag{12}$$

This case is a six dimensional supercomplex space-time. Since the supercomplex is a ring, it does not obey the commutative law of multiplication, i.e., $T_1T_2 \neq T_2T_1$. The three-dimensional time possesses a vector characteristic. This shows that T is q number in quantum theory, and has the irreversibility on the sequence of acting time.

If *i,j* and *k* are three coordinate axis, T is a three dimensional time, in which the simultaneity is represented by an equipotential surface $T(t_1,t_2,t_3) = C$. In this case, time possesses character of vector. Derivative of a function *f* for time corresponds the directional derivative:

$$\frac{\partial f}{\partial T} = \frac{\partial f}{\partial t_1}\cos\alpha + \frac{\partial f}{\partial t_2}\cos\beta + \frac{\partial f}{\partial t_3}\cos\gamma, \tag{13}$$

whose maximum is a graduation for time:

$$gradf = \frac{\partial f}{\partial t_1}\vec{i} + \frac{\partial f}{\partial t_2}\vec{j} + \frac{\partial f}{\partial t_3}\vec{k}. \tag{14}$$

This is a vector, which is perpendicular to the equipotential surface T=C, and points to a direction of function increase. We suppose that it may describe arrow of time, and corresponding the irreversibility of time. If the function is entropy, the graduation may define direction of time. Moreover, we may define a divergence of time $T(t_1,t_2,t_3)$:

$$divT = \frac{\partial t_1}{\partial x} + \frac{\partial t_2}{\partial y} + \frac{\partial t_3}{\partial z}. \tag{15}$$

This seems to represent that high-dimensional time may be divergent. We may also define a rotation of time:

$$rotT = (\frac{\partial t_3}{\partial y} - \frac{\partial t_2}{\partial z})\vec{i} + (\frac{\partial t_1}{\partial z} - \frac{\partial t_3}{\partial x})\vec{j} + (\frac{\partial t_2}{\partial x} - \frac{\partial t_1}{\partial y})\vec{k}. \tag{16}$$

This connects possibly to that high dimensional time may be spiral development.

At present various higher dimensional theories are increase of space. We have increased dimension of time to three [5]. If space-time theory combines biquaternion, time will be able to higher dimensional. It is consistent with the Kaluza-Klein theory and superstring. But, these dimensions between space and time may be re-allocated and developed for higher dimensional supersymmetry theory, superstring and various unified theories. For example, we derived some new representations of the supersymmetric transformations, and introduced the supermultiplets. So various formulations (includes equations, commutation relations, propagators, Jacobi identities, etc.) of bosons and fermions may be unified [15]. Such the mathematical characteristic of particles is proposed: bosons correspond to real number, and fermions correspond to imaginary number, respectively. Fermions of even (or odd) number form bosons (or fermions), which is just consistent with a relation between imaginary and real number. The imaginary number is only included in the equations, forms, and matrixes of fermions. This is connected with relativity. The unified forms of supersymmetry are also connected with the statistics unifying Bose-Einstein and Fermi-Dirac statistics [16]. Therefore, a possible direction of development is the higher dimensional complex space.

Further, in the higher dimensional time we introduced similarly a fractal time:

$$T_D = n_t + d_t, \tag{17}$$

which may be 2.3 or 0.7 dimensional time, etc. This may be developed to the complex time whose dimension is changeable, even supercomplex dimensional time. Such space and time possess completely symmetry. Fractal possesses the self-similarity of structure, and similarly the fractal time possesses biological Haeckel re-evolution law.

The higher dimensional and fractal time introduced will derive various developments of different theories. For example, these fourth components which correspond time in the four-dimensional vectors, energy, frequency and density, etc., may extend to the higher dimensional and fractal. In a word, it may be constructed preliminarily that the higher dimensional, fractal, complex and supercomplex space-time theory covers all [5].

L.Nottale studied fractal space-time and a principle of scale relativity, according to which the laws of physics must be such that they apply to coordinate systems whatever their state of scale, whose mathematical translation is the requirement of scale covariance of the equations of physics [17].

## 4. Generalized Noether Theorem and New Research of Space-Time

In nature there are various arrows of time [18,19], which are namely irreversibility of time. The Noether theorem connects closely space-time, and pointed out: if a system under a certain transformation group is invariance, the symmetry will produce necessarily a certain conservative quantity. For the continuous symmetric groups, any system, which may be represented by Lagrangian *L*, corresponds to a continuous equation,

$$\partial_\mu f_\mu = 0. \tag{18}$$

Here the four-dimensional vector

$$f_\mu = (L\delta_{\mu\nu} - \frac{\partial L}{\partial \partial_\mu \psi_\alpha} \partial_\nu \psi_\alpha)\delta x_\nu + \frac{\partial L}{\partial \partial_\mu \psi_\alpha} \delta \psi_\alpha, \tag{19}$$

and its integral are conservative. A well-known example is that uniform space-time corresponds to conservation of momentum-energy.

We develop the Noether theorem to a generalized formulation:

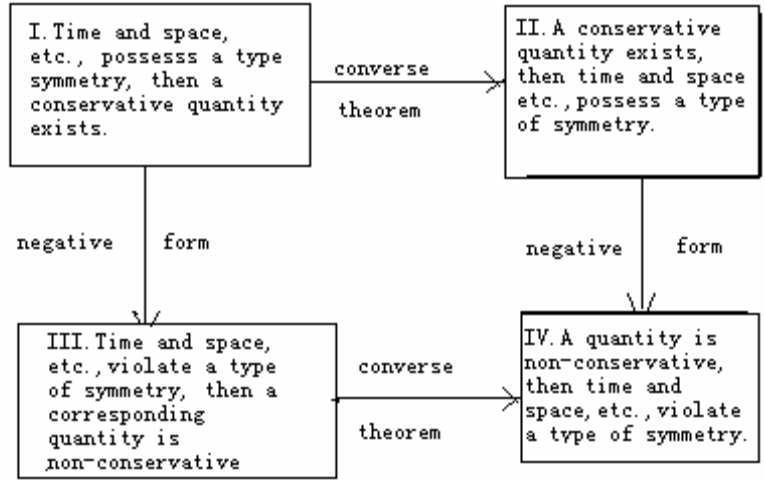

A particular case of the above results is that a type of symmetry is direction of time. Such irreversibility of time should correspond to non-conservation of certain quantity, for example, entropy. The simplest example is that when change of energy has a direction, the arrow of time may be defined from this. Further, all change exists in time, so change of any quantity with the direction may be applied to define the arrow of time.

Because the invariance under time reflection cannot obtain a constant, it cannot also obtain a definite non-conservative quantity when time reflection does not hold. The irreversibility of time in different regions seems to be due to different non-conservative quantities in these cases. Such different arrows of time are produced.

Recently, Hurth and Skenderis researched quantum Noether method [20]. In above statement, we discuss irreversibility of time in a high dimensional time. Here we propose another universal mathematical form on arrow of time: Assume that a certain quantity $f_\mu$ is non-conservative, so

$$\partial_\mu f_\mu = Q , \qquad (20)$$

where $Q$ is a dissipative term. In the field theory $\partial_\mu f_\mu$ represents $\mu$-dimensional divergence. According to the $\mu$-dimensional Gauss law,

$$\iint \cdots \int_{\mu-\text{dim.}} \partial_\mu f_\mu dV = \oiint \cdots \oiint_{(\mu-1)-\text{dim.}} f_\mu dS = \iint \cdots \iint_{\mu-\text{dim.}} QdV . \qquad (21)$$

For a conservation process a flux through a closed ($\mu$-1)-dimensional "curved surfac" is zero. While for the non-conservation process the flux is not zero, i.e., $f_\mu$ is changeable. This implies that a necessary condition of a conservation quantity (including conservation of energy) is an isolated system. Further, we think that the two necessary conditions for the conservation of energy and for reversibility of time should include: (a). Any frictional forces must be neglected; (b). System is isolated, which is stated deeply by the theory of the dissipative structure. The condition (b) is inattentive generally. Isolated systems and corresponding reversibility of time are a few idealized cases in macroscopic world. But, (b) is necessary. Because certain flow exists for any open system, from this a direction may be defined. A trajectory is indeed reversible for time, but input and output are two different directions. Therefore, nature is almost irreversible.

Reversibility of time corresponds to existence of inverse element, and irreversibility of time corresponds to a Markov process and semigroup. A factor $e^{at}$ or a generalized evolutionary operator $U_t = e^{Lt}$ defines a dissipative semigroup. Combining some known results, we proposed a fundamental operator [4,21],

$$p_\mu = -i\hbar(F\frac{\partial}{\partial x_\mu} + i\Gamma_\mu) , \qquad (22)$$

where $F$ and $\Gamma_\mu$ are corrected factor and additive term, respectively, and both may be nonlinear forms. From a corresponding equation

$$HU = -i\hbar(F_0\frac{\partial}{\partial t} + i\Gamma_0)U , \qquad (23)$$

or quantum theory, a form of $U_t$ may be solved when $F_0, \Gamma_0$ and H are not obvious functions of time.

The most known-well arrow of time is the second law of thermodynamics and increasing entropy. But, it is based on statistical independence, etc. We proposed that if interactions and fluctuations exist among various subsystems of an isolated system, these prerequisites will not hold. Entropy is not an additive quantity, and the decrease of entropy on the isolated system is possible [22]. Since fluctuations can be magnified due to internal interactions under a certain condition, the equal-probability does not hold, and entropy would be defined as $S(t) = -k\sum_r P_r(t)\ln P_r(t)$. From this we calculated decrease of entropy in a special internal condensed process [23]. Internal interactions, which bring about inapplicability of the statistical independence, cause possibly decreases of entropy in an isolated system. This possibility is researched for attractive process, internal energy, system entropy, and nonlinear interactions, etc [23]. In these cases, the statistics and the second law of thermodynamics should be different. An isolated system may form a self-organized structure, whose entropy is smaller, can be formed. Further, we find negative temperature is contradiction with usual meaning of temperature and with some basic concepts of physics and mathematics. Negative temperature is based on the Kelvin scale and the condition dU>0 and dS<0. Conversely, there is also negative temperature for dU<0 and dS>0. It will derive necessarily decrease of entropy [24].

When decrease of entropy is possible, not only the arrow of time on entropy will be an opposite direction, and general reversibility of time will be resumed. Further, PCT invariance and some evolutions of many systems from particle and biology to star and cosmology will open up new prospects.

In the special relativity, energy is $E = m_0 c^2 / \sqrt{1-(v/c)^2}$, the time interval at the same position is $dt = dt_0 / \sqrt{1-(v/c)^2}$, and space interval at the same instant is $dl = dl_0\sqrt{1-(v/c)^2}$. From this we derive necessarily a result: energy is in direct proportion to the time-interval at the same position in space, and is in inverse proportion to the space-interval at the same instant of time. The result may extend to the general relativity, which is also a development of space-time that depends to mass and its move. Combining the de Broglie relation $v\bar{v} = c^2$ in quantum theory, the same conclusion may be obtained, and these quantitative relations are [4]:

$$E = F_t \frac{hc^2}{l_0^2} dt \quad \text{and} \quad E = F_l \frac{hc^2}{v_0} \frac{1}{dl}. \tag{24}$$

Energy is quantum, so space and time are also quantum. Let

$$h\nu = F_t \frac{hc^2}{l_0^2} dt \quad \text{and} \quad h\nu = F_l \frac{hc^2}{v_0} \frac{1}{dl}, \tag{25}$$

the minimum space interval (space quantum) and the minimum time interval (time quantum) which correspond to quantum energy are respectively:

$$dl_0 = \frac{F_l c^2}{v_0 \nu} \quad \text{and} \quad dt_0 = \frac{l_0^2 \nu}{F_t c^2}. \tag{26}$$

If both are the Planck space scale $l_P = \sqrt{G\hbar/c^3} \approx 1.6 \times 10^{-33}$ cm and the Planck time scale $t_P = l_P/c \approx 0.5 \times 10^{-43}$ sec, and assume $l_0 = l_P, v_0 = c$, there will be a unifying $F_t = F_l = t_P \nu$. We think that this is a development of Noether theorem. The space is quantum structure for the string theory and the loop quantum gravity.

Further, it is consistent with the space-time uncertainty relation $\Delta X \Delta T \geq l_s^2$, which is proposed by Yoneya in 1989 based on the typical spatial extension $\Delta X \sim l_s^2 E$ of strings with energy E. This implies the simple relation for the indeterminacies of the space and time lengths. It can be derived as a direct consequence of the world-sheet conformal invariance. Li and Yoneya discussed that the space-time uncertainty relation of the form $\Delta X \Delta T \gtrsim \alpha'$ for the observability of the distances with respect to time is universally valid in string theory including D-branes. This relation, combined with the usual quantum mechanical uncertainty principle, explains the key qualitative features of D-particle dynamics [25]. Moreover, there is the Bekenstein's bound for information.

In electromagnetic field, the energy

$$E = m_0 c + e\varphi = m_0 c(1 + e\varphi/m_0 c) \tag{27}$$

derives a curved space-time. But, the second corrected term holds only for charged bodies, and correction is the same for the same ratio of charge to mass e/m. It is namely the principle of equivalence for the electromagnetic field [26]. This proves a possibility on the existence of electromagnetic space-time. Combining the above conclusion, the electromagnetic time should be:

$$dt = dt^0 (1 + e\varphi/m_0 c). \tag{28}$$

The most basic energy-mass and space-time in all nature are connected quantitatively. They construct a unified background for various scientific descriptions.